\documentclass[letterpaper,twocolumn,10pt]{article}
\usepackage{amsthm}
\usepackage{usenix2019,epsfig}
\usepackage{url}
\usepackage{booktabs}
\usepackage{threeparttable}
\usepackage{siunitx}
\usepackage{graphicx}
\usepackage{xcolor}
\usepackage{wasysym}
\usepackage{makecell}
\usepackage{flushend}
\graphicspath{ {./images/} }

\usepackage{array, tabularx}
\usepackage[hidelinks]{hyperref}

\newenvironment{conditions*}
  {\par\vspace{0.5\abovedisplayskip}\noindent
   \tabularx{\columnwidth}{>{$}l<{$} @{${}={}$} >{\raggedright\arraybackslash}X}}
  {\endtabularx\par\vspace{0.5\belowdisplayskip}}

\newcommand{\para}[1]{\vspace{0.08in}\noindent\textbf{#1 }}

\newcommand*\rot{\rotatebox{90}}

\newcommand{\squishlist}{
 \begin{list}{$\bullet$}
  { \setlength{\itemsep}{0pt}https://www.overleaf.com/6476192976jtxbnpcbgwbz
    \setlength{\parsep}{1pt}
    \setlength{\topsep}{1pt}
    \setlength{\partopsep}{0pt}
  }
}
\newcommand{\squishend}{\end{list}}

\begin{document}

% Reduce gap around equation
\setlength{\abovedisplayskip}{0.5em}
\setlength{\belowdisplayskip}{0.5em}

%don't want date printed
\date{}

\title{The Price of Free Illegal Live Streaming Services}
%for single author (just remove % characters)
\author{%
{\rm Hudson Ayers}\\
Stanford University
\and
{\rm Luke Hsiao}\\
Stanford University
%Name Institution
% copy the following lines to add more authors
% \and
% {\rm Name}\\
%Name Institution
\vspace{-10em}} % end author

\maketitle

% Use the following at camera-ready time to suppress page numbers.
% Comment it out when you first submit the paper for review.
% \thispagestyle{empty}

\subsection*{Abstract}

As Internet streaming of live content has gained on traditional cable TV
viewership, we have also seen significant growth of free live streaming services
which illegally provide free access to copyrighted content over the Internet.
Some of these services draw millions of viewers each month. Moreover, this
viewership has continued to increase, despite the consistent coupling of this
free content with deceptive advertisements and user-hostile tracking.

In this paper, we explore the ecosystem of free illegal live streaming services
by collecting and examining the behavior of a large corpus of illegal sports
streaming websites. We explore and quantify evidence of user tracking via
third-party HTTP requests, cookies, and fingerprinting techniques on more than
$27,303$ unique video streams provided by $467$ unique illegal live streaming
domains. We compare the behavior of illegal live streaming services with
legitimate services and find that the illegal services go to much greater
lengths to track users than most legitimate services, and use more obscure
tracking services. Similarly, we find that moderated sites that aggregate links
to illegal live streaming content fail to moderate out sites that go to
significant lengths to track users. In addition, we perform several case studies
which highlight deceptive behavior and modern techniques used by some domains to
avoid detection, monetize traffic, or otherwise exploit their viewers.

Overall, we find that despite recent improvements in mechanisms for detecting
malicious browser extensions, ad-blocking, and browser warnings, users of free
illegal live streaming services are still exposed to deceptive ads, malicious
browser extensions, scams, and extensive tracking. We conclude with insights
into the ecosystem and recommendations for addressing the challenges highlighted
by this study.

\section{Introduction}
\label{sec:intro}

Countless studies in the past decade have examined misbehaving websites which
attempt to exploit users. Traffic to these websites is often considered as
driven by exploiting innocent users, whether via phishing/spam, URL
misdirection, or other social engineering techniques. As a result, many security
efforts towards protecting users from these websites has focused on a)
preventing users from ending up on these sites at all and b) on making it
difficult for sites to exploit users without the users' noticing, giving them
the chance to navigate away. However, such efforts fail to protect one group of
Internet users---those who are willing to tolerate abnormal or undesirable web
behavior because the service they seek would otherwise cost money. As a result,
websites which illegally provide copyrighted content for free are able to
present user-hostile, even blatantly malicious interfaces without fear of losing
users as a result. Combine the behavior of these websites with the large number
of users that visit them, and you have a phenomenon worth studying.

Of course, illegal live video streaming is not the only instance which meets the
aforementioned criteria. While online piracy takes many forms, it has been
traditionally dominated by peer-to-peer platforms like BitTorrent. Despite this,
streaming presents an interesting use case---the attack vector for BitTorrent
platforms is obvious, as users are required to download untrusted files outside
the browser, and there is little expectation of safety. Video streaming, on the
other hand, is often considered by users as a lower-risk alternative, where the
browser is assumed to provide protection from harm. This perceived lower-risk
has likely been a significant factor in the meteoric rise of illegal video
streaming, and is an assumption worth investigating.

Like legitimate live video streaming services, \emph{illegal} live streaming
services exist to profit their operators. Hundreds of illegal live media
streaming services and thousands of link-aggregating websites have emerged to
lure viewers with free content and consequently attempt to monetize their
ill-gained traffic through deceptive ads, malware, tracking, and other malicious
behavior~\cite{rafique2016free}. Users willingly to expose themselves to these
risks by clicking through warnings and intrusive overlay ads in order to
received paid content for free---even though this intrusive and malicious
behavior would discourage them from visiting legitimate websites. Many of these
sites receive millions of visits each month, which likely places them among the
most visited sites that deliberately exploit their users. The surfeit of users
that willingly expose themselves to malicious behavior on these websites puts
these sites in a uniquely abusable position.

Previous research has focused on understanding and classifying the sites which
aggregate links to these illegal streams~\cite{rafique2016free}, illegal
on-demand streaming cyberlockers~\cite{ibosiola2018movie}, or the security and
privacy of legitimate streaming sites~\cite{nikas2018know}. In this study, we
explore the ecosystem of free illegal streaming services with a specific focus
on services which host streams to \emph{live} broadcasts. These websites are
highly transient in nature: they may change domains frequently, intentionally
avoid being indexed by search engines, and only host paths to live streams for
the duration the event is live (redirecting to an innocuous page after an event
is over). Consequently, these sites often avoid the scrutiny of law enforcement
and researchers. To study illegal live streams, we use live sports streams as a
proxy. Sports media rights are estimated to be valued at $\$20.6$B by
2019~\cite{heitner2015sports}. Illegal live sports broadcasts are primarily
viewed live by millions of viewers each month, making it a reasonable
representative of the live streaming ecosystem. Further, given the quantity and
diversity of these sites, we hypothesize that the malicious behaviors of these
websites represents a decent survey of malicious tactics used on the Internet
today by websites without fear of regulatory retribution.

\para{Contributions} Our methodology for studying the free illegal live
streaming ecosystem consists of three main phases: (1) collecting links to
illegal live streams, (2) storing the source and recording the behavior of these
websites, and (3) analyzing this data for evidence of malicious behaviors. To
summarize our contributions, in this study, we present:

\begin{itemize}
  \item A contemporary snapshot of the free illegal live streaming ecosystem,
    showing that the number of channel providers has increased since
    \cite{rafique2016free} and that mainstream sites are used as aggregators
    (Section~\ref{sec:ecosystem}).
  \item A system for automatically detecting new links to illegal live streaming
    sites as they appear, and for subsequently crawling these sites to collect
    information on them (Section~\ref{sec:method}).
  \item Evidence that illegal streaming sites seek to track and identify users,
    suggesting that these entities may have profiles for their viewers
    (Section~\ref{sec:tracking}).
  \item A comparison of illegal and legal live stream sites illustrating how
    illegal sites are more interested in tracking users
    (Section~\ref{sec:legit}).
  \item Evidence that moderation and crowd-source voting are likely less
    effective than users expect at protecting them from malicious sites
    (Section~\ref{sec:moderation}).
  \item A collection of case studies for illegal live streaming websites which
    serve as a lens into how modern cybercriminals attempt to exploit naive
    users. (Section~\ref{sec:case_studies}).
\end{itemize}

The remainder of the paper is structured as follows. Section~\ref{sec:related}
highlights related work and frames our study in the context of previous
literature. Section~\ref{sec:ecosystem} presents the ecosystem of illegal live
streaming services. Sections~\ref{sec:method} and \ref{sec:collection} describe
our methodology for gathering and analyzing our dataset.
Section~\ref{sec:analysis} presents our results. Section~\ref{sec:case_studies}
dives into illustrative examples of illegal streaming sites. We discuss the
implications of our observations in Section~\ref{sec:implications}, provide
recommendations in Section~\ref{sec:recommendations}, and conclude in
Section~\ref{sec:conclusion}.

\section{Related Work}
\label{sec:related}

% Placed here just to ensure that it shows up on the correct page.
\begin{figure*}[!ht]
  \centering
  \includegraphics[width=1.0\textwidth]{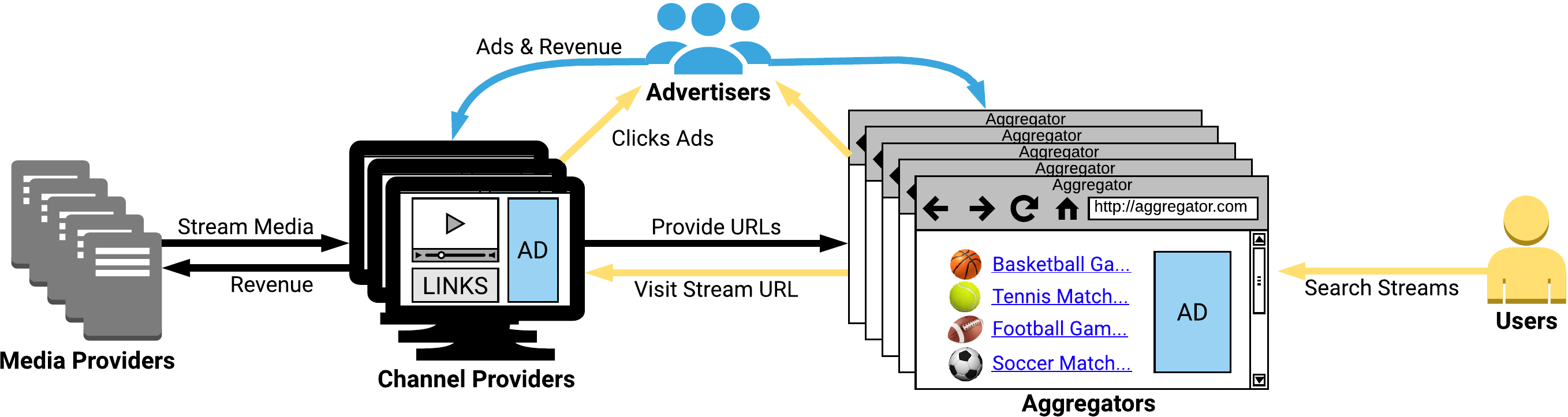}
  \caption{An overview of the ecosystem of illegal live streaming services. In
    some cases, \emph{media providers}, \emph{channel providers}, and/or
    \emph{aggregators} may be controlled by a single entity.
  }
	\label{fig:ecosystem}
\end{figure*}

\para{Web Security.} Many developments in web security have worked to minimize
the risk that malicious websites can present to unsuspecting users. Here, we
discuss three of these. First, tools like ad-blockers work to hide
advertisements on a site, which may reduce a user's exposure to deceptive
advertisements that link to scams or malware. Some deceptive ads can be detected
automatically~\cite{duman2014trueclick}. However, this is an arms race, as
anti-ad-blocking techniques have also improved~\cite{nithyanand2016adblocking,
iqbal2017ad}. Second, work on designing user experiences and warnings that
clearly alert users when their privacy may be at risk, such as security
warnings, have been shown to be effective in practice~\cite{akhawe2013alice,
felt2015improving}. Third, browser sandboxing has been employed to ensure that
access by websites to vulnerable aspects of the system is limited by specific
browser UI, such as directing all downloads to a specific directory, limiting
read access to files to certain OS-provided contexts, and presenting JavaScript
alerts outside of the rendering engine~\cite{chromium2008}. However, online
piracy presents a unique scenario in which users are often willing to ignore and
bypass these security warnings in order to obtain paid content for free.

\para{Measuring Online Tracking.} \cite{englehardt2016online} presents extensive
measurements of online tracking across the Alexa top million websites and
presents OpenWPM, the tool we utilize in our work to collect data on illegal
stream URLs. Similarly, \cite{binns2018measure} studies third-party tracking on
websites and mobile applications while \cite{fruchter2015} examines the
differences in tracking activity between geographic locations. While these
studies measure tracking on the web generally, they do not differentiate between
sites, and none focuses specifically on sites for which visiting these sites
could be considered criminal activity.

\para{Illegal Media Streaming.} \cite{nikas2018know} studies security and
privacy concerns related to on-demand media streaming services and targets
platforms that are known to host illegal content. Specifically, they study over
20 media streaming platforms (e.g., Kodi, Enigma 2, MediaTomb, etc.) and their
attack surfaces, and find that there are over $100,000$ devices using these
platforms which are discoverable through simple search queries. Similarly,
\cite{ibosiola2018movie} explores the ecosystem of illegal streaming from the
perspective of video piracy, where content is streamed on-demand, as opposed to
our work, which will focus specifically on live-streamed content.
\cite{leporini2015architectures} studies the architectures and protocols used to
stream illegal content over the Internet and explores the value chain from
content acquisition, preparation and distribution, web hosting, and content
discovery. This study considers peer-to-peer streaming as well as web streaming,
but does not study malicious behavior outside of breaking copyright law.

\para{Illegal Live Media Streaming.} \cite{rafique2016free} studies the
ecosystem of free live streaming websites with an analysis of over 5600
live-streaming domains discovered from live-streaming domains through aggregator
websites. This study does not focus on user tracking, and instead highlights
other aspects of their behavior such as trademark infringements, malware
distribution, and anti-ad-block techniques, and uses these intuitions to build a
classifier of these sites. Though this study is over two years old, it notably
does not include Reddit as one of their aggregators, despite our finding that
Reddit is now one of the most popular aggregators (see
Section~\ref{sec:method}).

\section{The Illegal Live Streaming Ecosystem}
\label{sec:ecosystem}

In the ecosystem of free live streaming services, there are five main parties
involved. The relationships between these parties are illustrated in
Figure~\ref{fig:ecosystem} and described below.

\para{Media Providers} acquire and stream media content. A media provider may be
a single individual streamer sending a video stream using software like
OBS\footnote{Open Broadcaster Software (\url{https://obsproject.com/})}, or a
large entity such as a broadcasting station. In the context of illegal live
streams, an example of a media provider is an individual who has a subscription
to a paid service, and rebroadcasts this content for free in real time.

\para{Channel Providers (CPs)} provide the infrastructure for live-streaming
media through a media server, which receives live video streams from media
providers and serves the streams to users. These entities host the webpages on
which streams can be viewed. As such, the webpages of channel providers are
where users ultimately land when searching for and viewing illegal live streams.
In the context of legitimate websites, Twitch and YouTube are examples of
channel providers. Examples of illegal services include sites like
\texttt{buffstreamz.com} or \texttt{watchsport.fun}~\footnote{Author's note: The
ads displayed on websites analyzed in this study vary frequently and widely, and
any of the sites referenced in this paper may at times display ads for adult
content. Please take appropriate caution if you choose to visit them.}.

\para{Aggregators} collect links to a variety of channel providers to allow
users to discover and browse streams that are available to view. Some
aggregators may offer search functionality (e.g.\ Reddit) while others simply
offer a list of streams which are available for viewing at the current point in
time (e.g.\ \texttt{firstrowonly.eu}). Some aggregators use crowd-sourcing to
gather URLs, while others may be run by a single individual or automated
service. In some cases, aggregators themselves may also serve as channel
providers, such that when a user clicks on a link to a live event, they do not
leave the aggregator domain.

\para{Advertisers} allow media providers, channel providers, and aggregators to
monetize their services through ads and overlays. Advertisers may be individuals
or ad networks, and play a key role in this ecosystem. Typically, aggregators
and channel providers will run JavaScript code which fetches advertisements from
an ad network. These ads may be banner ads, or more intrusive overlay ads that a
user is compelled to click in order to view a live stream, and typically open a
new tab that is redirected for tracking before ending on an advertisement page.
These advertisement pages are often deceptive or blatantly malicious, with many
offering benefits in exchange for the user installing a browser extension.

\para{Users} are the individuals watching these free illegal live streams on
these potentially malicious websites.

\section{Methodology}
\label{sec:method}

Our investigation is organized into three phases.

\para{Phase 1: Collection.} In Phase 1, we identify a list of popular
aggregators. Because we are most interested in understanding the behavior of
channel providers (who host the video pages users ultimately land on when
watching a live stream), we select eight popular aggregators (based on their
Alexa rankings) found in search engines, along with Reddit.

Notably, adding Reddit represents a departure from prior research on this topic.
Reddit is a top 20 website globally and is hosted in the United States, making
it substantially different from the aggregators described
in~\cite{rafique2016free}. Further, Reddit is different from other social media
platforms such as Facebook and Twitter in that the subreddits via which these
links are available are public, where sharing is not dependent on friend
networks, and where viewing links does not require users be a member of the
Reddit platform. We manually identify 14 subreddits which focus exclusively on
aggregating links to illegal sports streams. A comprehensive analysis of Alexa
and SimilarWeb data on Reddit indicates that Reddit receives between 16 million
and 85 million visits a month \textit{just from users looking for sports
streams}.

We monitor these aggregators for five weeks (11/2018 to 12/2018), to build a
database of channel providers linked to by these aggregators\footnote{Our
collection code is available at
\url{https://github.com/hudson-ayers/safe-sports-streams}.}. Our collection code
scraped each of these aggregators every 15 minutes for this five week period.

\para{Phase 2: Crawling.} In Phase 2, we automatically crawl these websites
using our own fork of OpenWPM v0.8.0~\cite{englehardt2016online}, which is built
upon Firefox and Selenium. OpenWPM allows us to collect data about a webpage
such as cookies, HTTP requests, and JavaScript calls. By default, OpenWPM simply
visits a webpage and collects information about the content that is immediately
available. In our experience, however, we find that many of these websites
redirect to what appear to be more malicious pages only when a user attempts to
click on the ``play'' button of the video. As a result, we extend OpenWPM with
commands for clicking the webpage and collecting data about the resulting
behavior. Due to time constraints, we do not crawl all of the URL we harvested
in Phase 1. Instead, we take a sample of the most recently harvested URLs from
each channel provider in order to ensure each is represented.

\para{Phase 3: Analysis.} Finally, in Phase 3, we analyze the behavior of the
visited webpages in the context of tracking. To understand the tracking behavior
of these websites, we leverage
EasyPrivacy\footnote{\url{https://easylist.to/}}---a filter list which seeks to
block web tracking. We compare the domains of third party HTTP requests and
cookies against EasyPrivacy to determine which of the requests and cookies set
by a webpage are associated with web tracking. Importantly, this metric likely
underestimates the amount of tracking, since EasyPrivacy is not exhaustive and
may fail to flag some resources as tracking~\cite{wills2016ad}.

Next, to understand the fingerprinting behavior of these sites, we implement
checks for canvas, font, and WebRTC fingerprinting using the techniques
described in \cite{englehardt2018thesis}. Canvas fingerprinting is detected by a
series of checks: (1) the canvas element must be greater than 16px in
\texttt{height} and \texttt{width}, (2) the JavaScript on the page must not call
\texttt{save}, \texttt{restore}, or \texttt{addEventListener}, (3) the
JavaScript must attempt to save an image with \texttt{toDataURL} or
\texttt{getImageData} with a size greater than 16px $\times$ 16px. Font
fingerprinting is flagged when a site calls \texttt{measureText} 50 or more
times. WebRTC fingerprinting is detected by whether or not the page accesses the
\texttt{localDescription} property of a \texttt{RTCPeerConnection}, as this
property covers all possible IP address retrievals. We then use the same
techniques for quantifying user tracking on legitimate sites in order to compare
the behavior of illegal and legitimate streaming sites\footnote{Our crawling and
analysis code is available at
\url{https://github.com/lukehsiao/price-of-free-illegal-streams}.}.

\subsection{Tracking Score}
We present a general metric for how malicious a particular channel provider
is based on the tracking techniques they utilize. This metric follows:
\begin{equation}
     \texttt{score} = 0.5*r + 3*c + 5*f
\end{equation}
where:
\begin{conditions*}
r     &  Ave. \# of HTTP requests to tracking domains for each visited channel
         provider URL. \\
c     &  Ave. \# of cookies set by tracking domains for each visited channel
         provider URL \\
f     &  \# of unique of fingerprinting methods used by the channel Provider.
         (Ranges from 0 to 3, i.e. a site using both canvas fingerprinting and
         font fingerprinting would have $c = 2$)
\end{conditions*}
\noindent We evaluate this metric for each channel provider. The higher the
score, the more invasive a site is. We acknowledge that these constants are
relatively arbitrary. Our basic reasoning is as follows: fingerprinting is the
hardest tracking mechanism to evade~\cite{acar2014web}, and thus is given the
highest multiplier. For tracking cookies and requests, we observe that cookies
are set less frequently, and scale them accordingly.

\subsection{Limitations}

First, many channel providers are protected by Cloudflare. Although there are
techniques to discover the true IP addresses through information
leakage~\cite{nixon2013ddos}, we did not implement these checks. Consequently,
our study does not include detailed analysis about the true geolocation of the
parties involved in the illegal live streaming ecosystem.

Second, due to time constraints, we do not perform a measurement of malware
distributed by these websites. OpenWPM does not include mechanisms for
collecting malicious downloads or browser extensions. We believe this is an
interesting area of future work and highlight some of the malicious extensions
served by these websites in Section~\ref{sec:case_studies}. Future work could
also seek to detect drive-by-downloads~\cite{kiire2016detecting}.

\subsection{Ethical Considerations}

The purveyors of the illegal sites we study are breaking U.S.\ copyright law,
and this paper discourages illegal streaming. We aim to observe a set of
websites which large numbers of users willingly and regularly visit despite
circumstantial evidence of malware and deceptive practices. Our methodology
benefits these illegal sites very little---while we may slightly increase ad
revenue in the process of our study, our visit numbers are insignificant
compared to the number of users these sites attract daily. We are not personally
taking revenue from the copyright holders of this content, as the majority of
our crawls occurred through headless browsers that do not display any video.
Finally, any added views to these websites from our study is likely offset by
any increased awareness of the hostile behavior that this study highlights.

\section{Collection Results}
\label{sec:collection}

As discussed in Section~\ref{sec:method}, we scrape stream URLs from several
manually selected popular aggregator sites. The aggregators that we target are
listed in Table~\ref{tab:aggregators}.

\begin{table}
  \scriptsize
  \centering
  \caption{Summary of Aggregator Sites}\label{tab:aggregators}
  \begin{threeparttable}
    \begin{tabular}{lrr}
    \toprule
    \textbf{Aggregator} & \textbf{Alexa Rank} & \textbf{Monthly Visits (100k)} \\
    \midrule
    \url{https://www.reddit.com}\tnote{$\ddagger$} & 18     & 16,000\tnote{$\dagger$} \\
    \url{http://livetv.sx}                         & 1,380  & 8,100                   \\
    \url{http://rojadirecta.me}                    & 8,755  & 2,000                   \\
    \url{https://www.stream2watch.org}             & 9,765  & 2,200                   \\
    \url{http://cricsports.sc}                     & 12,014 & 2,200                   \\
    \url{http://www.fromhot.com}                   & 21,680 & 170                     \\
    \url{http://firstrowonly.eu}                   & 27,001 & 1,100                   \\
    \url{https://www.mamahd.org}                   & 29,523 & 350                     \\
    \url{https://www.viprow.net}                   & 35,203 & 760                     \\
    \bottomrule
    \end{tabular}
    \begin{tablenotes}
      \item[$\dagger$] Approximation of monthly traffic exclusively to live
        sports stream subreddits based on search traffic leading to these
        subreddits: 16-85 million views.
      \item[$\ddagger$] Specifically, we look at the following subreddits:
        \texttt{/r/\{boxing, cfb, cricket, mlb, mma, motorsports, nba,
        ncaabball, nfl, nhl, putt, rugby, soccer, wwe\}streams}.
    \end{tablenotes}
  \end{threeparttable}
  \vspace{-2em}
\end{table}

\begin{table}
  \scriptsize
  \centering
  \caption{Top 10 Hosts for Channel Providers}\label{tab:channel}
  \begin{tabular}{llrrS[table-format=3.2]}
  \toprule
  \textbf{Hosting Company} & \textbf{Host Country} & \textbf{AS \#} & \textbf{\# CP} & \textbf{\% Streams} \\
  \midrule
  Cloudflare     & US & 13335  & 160 & 53.46 \\
  Google         & US & 15169  & 80  & 0.94  \\
  HostPalace     & NL & 134512 & 20  & 0.55  \\
  NameCheap      & US & 22612  & 19  & 0.20  \\
  MAROSNET       & RU & 48666  & 18  & 0.94  \\
  Quasi Networks & SC & 29073  & 11  & 23.17 \\
  DigitalOcean   & US & 14061  & 10  & 0.22  \\
  CompuByte      & GB & 39572  & 9   & 0.01  \\
  Lala Bhoola    & SE & 49453  & 9   & 0.36  \\
  BlueAngelHost  & BG & 206349 & 9   & 0.22  \\
  \bottomrule
  \end{tabular}
  \vspace{-1em}
\end{table}

\begin{figure}[t]
  \centering
  \includegraphics[width=1.\columnwidth]{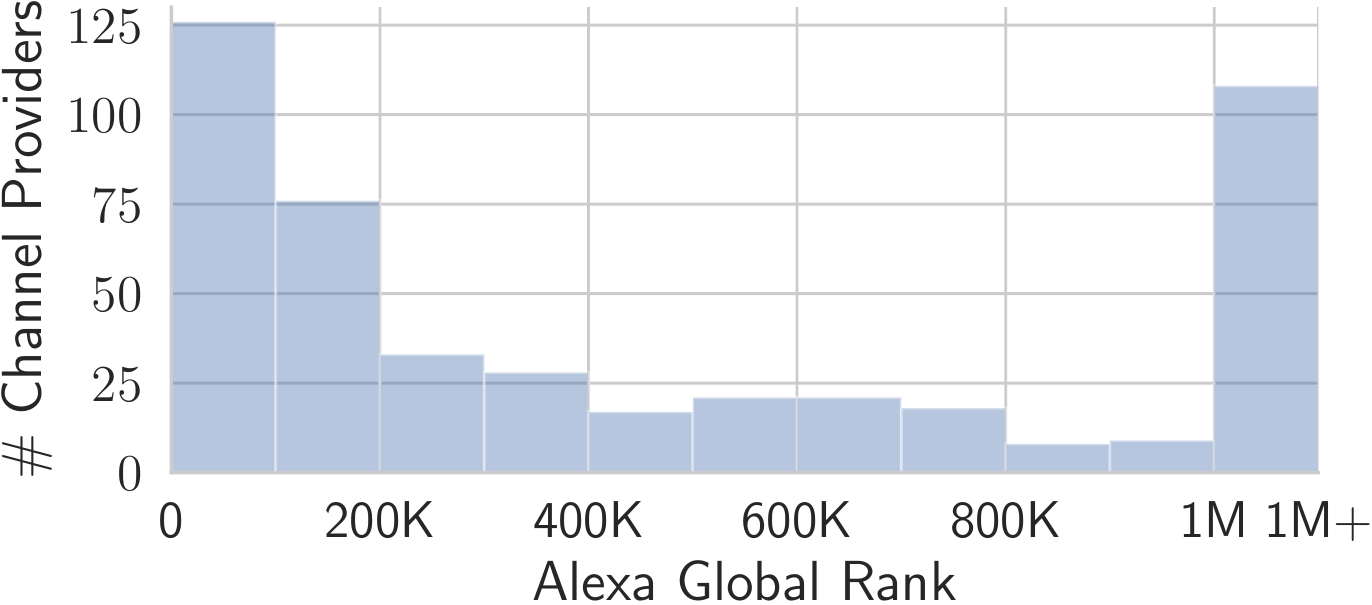}
  \caption{Distribution of Channel Providers by Alexa Rank.}
  \label{fig:alexa}
  \vspace{-1em}
\end{figure}

Through these aggregators, we collect $151,661$ unique URLs, spread across $467$
unique channel providers. Of these, we could not identify a host for $2$. Of the
remaining $465$ channel providers, we found that $160$ ($34\%$) were protected
by Cloudflare, which hides their true IP addresses. This data is summarized in
Table~\ref{tab:channel} where $74\%$ of our channel providers are hosted by just
$10$ hosts. Figure~\ref{fig:alexa} shows the global ranking distribution of the
channel providers---$111$ channel providers did not make the top million. The
top two Alexa-ranked channel providers were \texttt{youtube.com} and
\texttt{twitch.tv}---some streamers manage to sneak live sports content past the
filtering on these websites using misleading names. Sites such as
\texttt{buffstreamz.com} and \texttt{streamlabs.com} were ranked about 2000
globally, with \texttt{buffstreamz.com} ranked 344 in the U.S., ahead of
\texttt{nhl.com} and close to \texttt{nba.com} (the primary domains for
\textit{legally} streaming NHL and NBA games).

\section{Crawling \& Analysis Results}
\label{sec:analysis}

We perform three analyses to illustrate the behavior of illegal live streaming
sites. We first measure user tracking with respect to HTTP requests and cookies
from known-tracking domains as well as common fingerprinting techniques. Next,
we perform the same measurement on popular, legitimate sports streaming websites
in order to compare their behavior with illegal streaming sites. Finally, we use
the tracking score introduced in Section~\ref{sec:method} to generally compare
user tracking across sites and to assess the efficacy of heavily moderated
aggregators in protecting users from tracking.

\subsection{Evidence of User Tracking}
\label{sec:tracking}

As discussed in Section~\ref{sec:method} we break user tracking down into three
categories: third-party HTTP requests, cookies, and fingerprinting. We measured
channel providers by the prevalence of each of these tracking techniques, and
for each technique we ranked all of the channel providers we found by using the
tracking score for each channel provider, as described in
Section~\ref{sec:method}. These illegal streaming websites expose users to a
significant number of requests and cookies of known trackers. For each channel
provider, we measure the total number of third-party HTTP requests that are made
on each page we visit, the number of requests which are flagged a tracking
request by EasyPrivacy, and compute an average over all pages for a particular
channel provider. We do the same for tracking cookies.

Next, we analyzed each webpage's JavaScript behavior to detect whether or not
they were trying to perform device fingerprinting based on a series of
heuristics (see Section~\ref{sec:method}). We summarize the number of channel
providers suspected of fingerprinting in Table~\ref{tab:cp_fingerprinting}. We
observe that the $15$ ($3.2$\%) channel providers, engage in all three forms of
fingerprinting. Similarly, $128$ ($27.4$\%) channel providers engage in both canvas
and WebRTC fingerprinting.

\begin{table}
  \scriptsize
  \centering
  \caption{Channel Providers using Fingerprinting}\label{tab:cp_fingerprinting}
  \begin{tabular}{lrr}
  \toprule
  \textbf{Technique} & \textbf{\# CPs} & \textbf{\% CPs} \\
  \midrule
  Canvas Fingerprinting & 237 & 50.75 \\
  WebRTC Fingerprinting & 164 & 35.12 \\
  Font Fingerprinting   & 23  & 4.93  \\
  \bottomrule
  \end{tabular}
  \vspace{-1em}
\end{table}

Table~\ref{tab:top_sites} displays the top 10 illegal channel providers ordered
by tracking score, along with the prevalence of third-party HTTP requests to
trackers, tracking cookies, and fingerprinting techniques used. We include
the same information for the legitimate sites we study in
Section~\ref{sec:legit}.

\begin{table}
  \scriptsize
  \centering
  \caption{Tracking Behavior of Top CPs by Tracking Score}\label{tab:top_sites}
  \addtolength{\tabcolsep}{-3pt}
  \begin{tabular}{lrrcccr}
     & \textbf{\thead{Ave. \# \\ Requests}} &
    \textbf{\thead{Ave. \# \\ Cookies}} & \textbf{\thead{\rot{Canvas}}} &
    \textbf{\thead{\rot{Font}}} & \textbf{\thead{\rot{WebRTC}}} &
    \textbf{\thead{Tracking \\ Score}}\\
    \toprule
    \textbf{Illegal Channel Providers} & & & & & & \\
    \midrule
    \texttt{www.totalsportek.com} & 152.33 & 4.00 & \CIRCLE &         &         & 93.17  \\
    \texttt{ko.rivosport.co}      & 93.60  & 0.00 & \CIRCLE &         & \CIRCLE & 56.80  \\
    \texttt{nowwatchtvlive.ws}    & 108.23 & 0.00 &         &         &         & 54.12  \\
    \texttt{in.xrivonet.info}     & 81.71  & 0.00 & \CIRCLE &         & \CIRCLE & 50.85  \\
    \texttt{rvo8z.blogspot.com}   & 72.00  & 0.00 & \CIRCLE &         & \CIRCLE & 46.00  \\
    \texttt{www.gatehd.com}       & 67.70  & 0.13 & \CIRCLE &         & \CIRCLE & 44.25  \\
    \texttt{www.linesports.club}  & 56.87  & 0.00 & \CIRCLE & \CIRCLE & \CIRCLE & 43.43  \\
    \texttt{streamhd247.com}      & 63.30  & 0.03 & \CIRCLE &         & \CIRCLE & 41.75  \\
    \texttt{www.time4tv.net}      & 70.20  & 0.10 & \CIRCLE &         &         & 40.40  \\
    \midrule
    \textbf{Legitimate Channel Providers} & & & & & & \\
    \midrule
    \texttt{www.nba.com/nbatv}          & 42.00 & 0.00 & \CIRCLE &  &  & 26.00 \\
    \texttt{www.cbssports.com/live/}    & 38.67 & 0.00 &         &  &  & 19.33 \\
    \texttt{www.espn.com/watch/}        & 32.00 & 0.00 &         &  &  & 16.00 \\
    \texttt{www.tntdrama.com/watchtnt/} & 14.67 & 0.00 & \CIRCLE &  &  & 12.33 \\
    \texttt{www.nhl.com/tv/}            & 20.33 & 0.00 &         &  &  & 10.17 \\
    \texttt{www.willow.tv/}             & 7.33  & 0.00 &         &  &  & 3.67  \\
    \texttt{www.foxsportsgo.com/}       & 7.00  & 0.00 &         &  &  & 3.50  \\
    \bottomrule
  \end{tabular}
  \addtolength{\tabcolsep}{3pt}
  \vspace{-1em}
\end{table}

\begin{table}
  \scriptsize
  \centering
  \caption{Top Trackers}\label{tab:top_trackers}
  \begin{tabular}{lr}
  \toprule
  \textbf{Trackers on Illegal Sites}    & \textbf{\% All Tracking} \\
  \midrule
  \texttt{www.google-analytics.com} & 27.84 \\
  \texttt{cm.g.doubleclick.net}     & 7.89  \\
  \texttt{trk.vidible.tv}           & 7.63  \\
  \texttt{stats.g.doubleclick.net}  & 4.14  \\
  \texttt{px.moatads.com}           & 3.79  \\
  \texttt{counter.yadro.ru}         & 2.92  \\
  \texttt{c.mgid.com}               & 2.56  \\
  \texttt{d3.c3.b1.a1.top.mail.ru}  & 2.43  \\
  \texttt{k.streamrail.com}         & 2.30  \\
  \texttt{mc.yandex.ru}             & 2.17  \\
  \midrule
  \textbf{Trackers on Legitimate Sites} & \textbf{\% All Tracking} \\
  \midrule
  \texttt{cm.g.doubleclick.net}         & 18.60 \\
  \texttt{analytics.twitter.com}        & 4.32  \\
  \texttt{ma61-r.analytics.edgekey.net} & 3.99  \\
  \texttt{ib.adnxs.com}                 & 3.65  \\
  \texttt{px.moatads.com}               & 3.65  \\
  \texttt{www.facebook.com}             & 2.99  \\
  \texttt{stats.g.doubleclick.net}      & 2.99  \\
  \texttt{ping.chartbeat.net}           & 2.99  \\
  \texttt{www.google-analytics.com}     & 2.99  \\
  \texttt{t.co/i/adsct}                 & 2.66  \\
  \bottomrule
  \end{tabular}
  \vspace{-1em}
\end{table}

\subsection{Comparison to Legitimate Sites}
\label{sec:legit}

Observing the frequency of tracking on these sites is only useful within the
context of the frequency of such tracking on other websites. To provide this
context, we repeated our measurement process for a small corpus of legitimate
websites via which users can pay to watch sports. Specifically, we selected the
services WatchESPN, NHLTV, CBS Sports, WatchTNT, NBATV, Fox Sports Go, and
Willow TV, because these are widely used services, and because they expose a
viewing page accessible to non-logged-in users.

The results can be found in Table~\ref{tab:top_sites}. A list of the most
frequently detected trackers exists in Table~\ref{tab:top_trackers}, alongside
the percentage of total tracking requests that each tracker was responsible for.
We find that the legitimate sites make significantly fewer requests to known
tracking domains than the worst illegal websites. We see that Google Analytics
and DoubleClick are the only top 10 trackers shared between illegal and
legitimate streaming services. We also observe that all of the trackers used by
legitimate sites are either \texttt{.com} or \texttt{.net} domains (the
\texttt{t.co} domain is owned by Twitter). On the illegal sites we find a much
more diverse set of trackers, often with obscure names, and with some domains
corresponding to countries outside the US (Russia and Tuvalu).

The biggest difference between the legal and illegal sites shows up when looking
at cookies and fingerprinting. For all legal streaming sites analyzed, only 2
total instances of fingerprinting behavior was observed, and none of the cookies
set by these sites were detected as tracking cookies according to EasyPrivacy.

\subsection{Efficacy of Moderation}
\label{sec:moderation}
\begin{figure}
  \centering
  \includegraphics[width=1.\columnwidth]{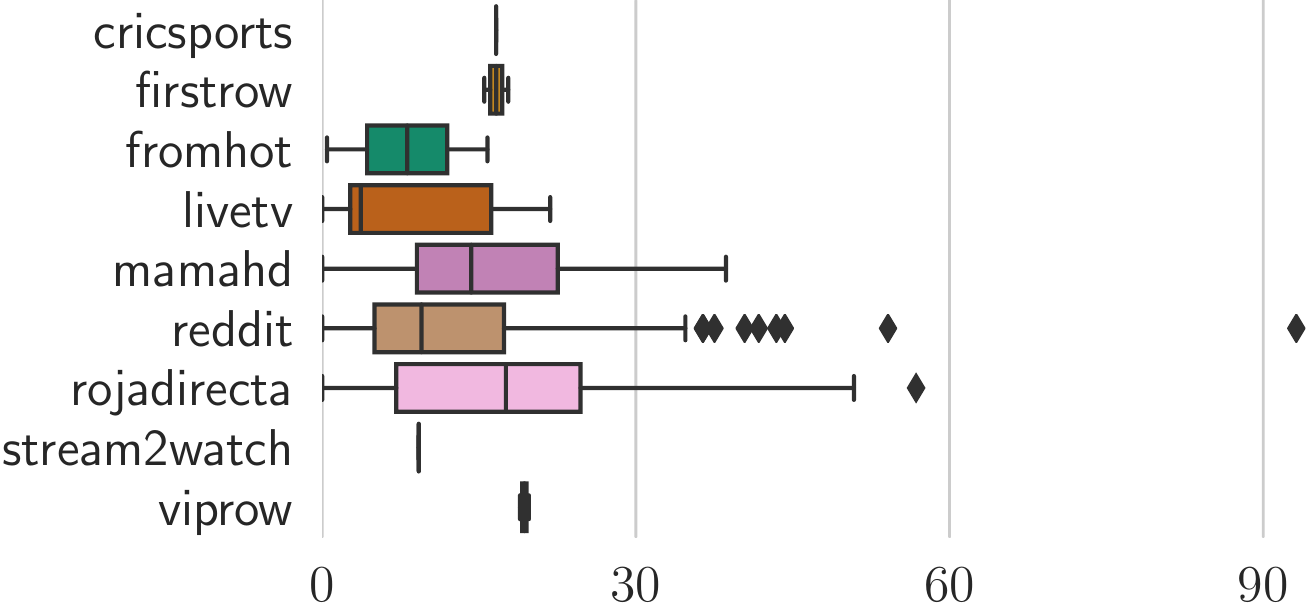}
  \caption{Tracking Score by Aggregator.}
  \label{fig:agg_privacy}
  \vspace{-1em}
\end{figure}

It is interesting to analyze trends across streaming providers and aggregators,
to answer questions such as ``Do moderated streaming aggregators do a better job
of filtering out the most malicious websites?''. Specifically, we focus on the
use of Reddit as a moderated aggregator of illegal streams. Reddit as an
aggregator has exploded in recent years, likely taking customers away from many
of the other unmoderated aggregators discussed in this paper. One reason for
this may be that viewers feel safer accessing streams linked to on subreddits
with moderators that promise to remove links which violate subreddit rules. In
fact, some subreddits go so far as to mark certain streamers as ``verified'',
meaning that they have been sharing their streams for an extended period of
time, have communicated personally with the moderators, and have not been outed
as abusing their users. As discussed in Section~\ref{sec:method}, we rely on our
tracking score metric to evaluate the effectiveness of moderation by averaging
the score of each channel provider across the aggregators that link them. The
results of this can be seen in figure~\ref{fig:agg_privacy}.

This figure reveals that the distribution of tracking scores reflects the nature
of the aggregator site. For example, Reddit is crowd-sourced and manually
moderated. Despite a high number of submissions of highly variable quality
(reflected in the large number of outliers), Reddit's moderation appears to help
it maintain an average tracking score that is lower than many of the other
aggregators. In contrast, \texttt{firstrowonly.eu} is not crowd sourced and acts
as both an aggregator and channel provider. It exhibits much less variance in
tracking scores, along with a higher average score, indicating more tracking
activity. Ultimately, this figure shows that Reddit does not successfully
moderate away the worst websites by this metric, suggesting that moderation is
more helpful for filtering out non-functioning streams than it is at removing
sites that exhibit malicious behavior.

\begin{figure}
  \centering
  \includegraphics[width=1.\columnwidth]{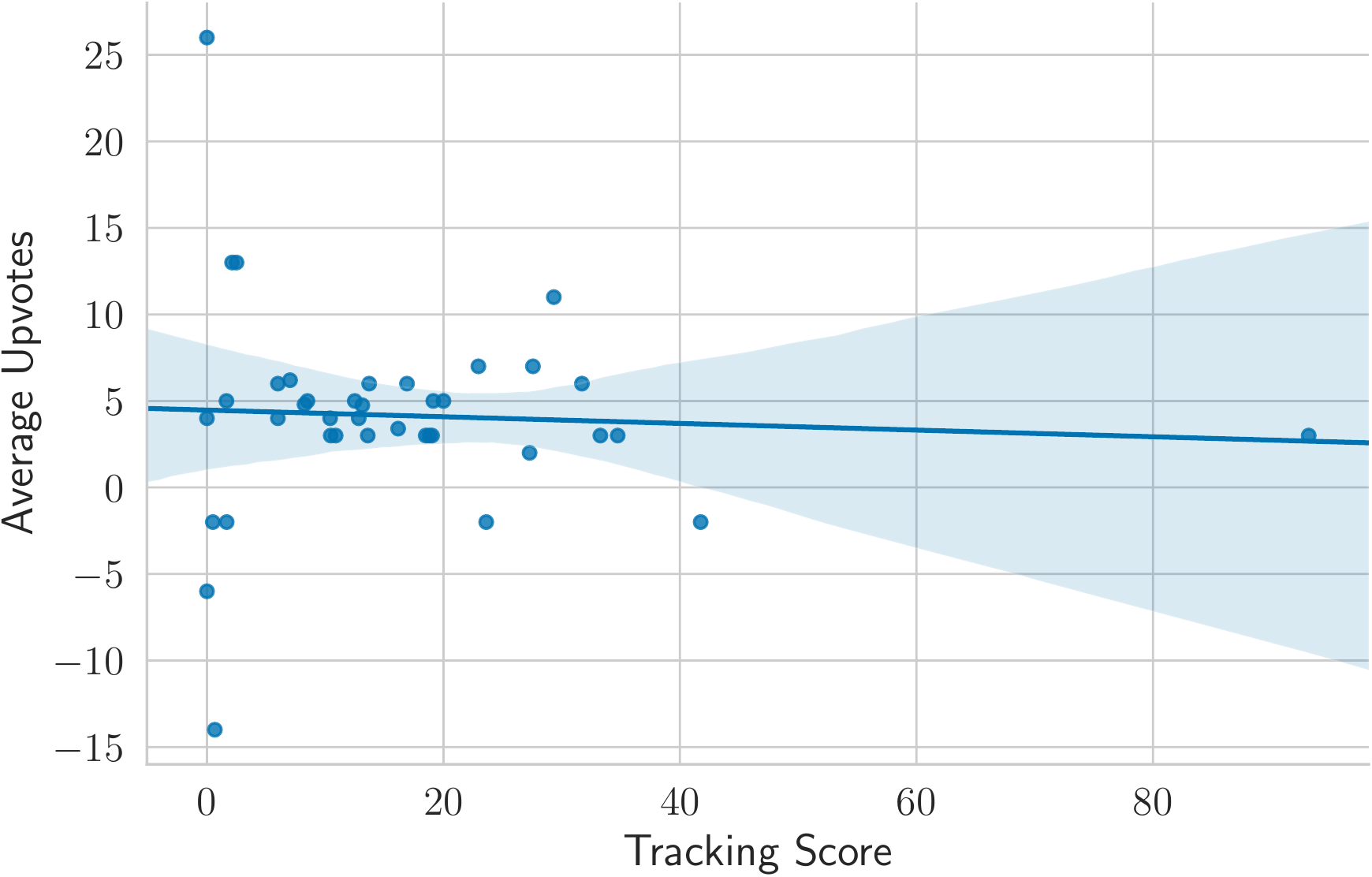}
  \caption{Tracking Score vs Upvotes on Reddit.}
  \label{fig:moderation}
  \vspace{-1em}
\end{figure}

One other interesting aspect of Reddit is that users are able to vote on the
streams posted to this site. We were curious as to whether this voting would
punish sites which track users, as we suspected that sites which track users
might also be more likely to display deceptive ads or other features which annoy
users. To do so, we plotted the average number of upvotes on each post which
linked to a channel provider from a streaming subreddit against the tracking
score of that channel provider. In order to disregard posts which had not been
voted on much, we did not count posts with 0, 1, or 2 upvotes. The results of
this can be seen in figure~\ref{fig:moderation}. As this figure shows, there is
minimal correlation between upvotes and tracking score. This suggests that user
opinions are not a good method for distinguishing malicious sites in this
ecosystem.

\section{Case Studies}
\label{sec:case_studies}

We manually investigated several channel providers and aggregators and noticed a
wide variety of interesting techniques and strategies which could be studied in
future work. We highlight several of the techniques we observed below.

\para{Malicious Extensions} As discussed in Section~\ref{sec:method},
limitations with OpenWPM made it infeasible for us to conduct a widespread study
of malware on these sites. Despite this, we did manually visit dozens of these
sites and clicked on ``download now'' and other similar links which we thought
were most likely to lead to malware (these links were typically accompanied by
some promise of helpful software or free media). Almost always, these links
would open a new webpage in another tab, and this new webpage would provide
instructions on how to install a specific Chrome extension, alongside some claim
that the extension was necessary to use whatever product the original link
promised. These extensions were accompanied by vague descriptions and few or no
reviews, and most were uploaded to the Chrome Webstore within the last couple of
months. These extensions often had the ability to read users entire browsing
history, and to read and modify data on all websites users might visit. Some
even had the ability to replace the page users see when they open a new tab and
manage users downloads folder. Notably, some of these extensions had tens or
hundreds of thousands of users. Limited existing reviews largely corroborated
our suspicion that these extensions were adware or spyware. A description of
these extensions can be found in Table~\ref{tab:extensions}.

\begin{table}
  \scriptsize
  \centering
  \caption{Suspected Malicious Chrome Extensions}\label{tab:extensions}
  \begin{threeparttable}
  \addtolength{\tabcolsep}{-2pt}
  \begin{tabular}{lrccccccc}
    \toprule
    \textbf{Extension Name} & \textbf{\# Users} & [1] & [2] & [3] & [4] & [5] & [6] & [7] \\
    \midrule
    \texttt{\href{https://chrome.google.com/webstore/detail/mytransitguide/npmoikddpdgbhgbkjgjemncoegpojpng}{MyTransitGuide}}                     & 2,759,523 &         & \CIRCLE & \CIRCLE & \CIRCLE &         & \CIRCLE & \CIRCLE \\
    \texttt{\href{https://chrome.google.com/webstore/detail/streamall/kfkkfgfmlliomakfjlijbpjniajpgboa}{StreamAll}}                               & 154,978   &         & \CIRCLE & \CIRCLE &         &         &         & \CIRCLE \\
    \texttt{\href{https://chrome.google.com/webstore/detail/pdf-maker/lpmedofaljioejmmdbbdbkadeenbgjcf}{PDF Maker}}                               & 124,072   & \CIRCLE &         &         &         & \CIRCLE &         &         \\
    \texttt{\href{https://chrome.google.com/webstore/detail/televisionfanatic/oahfdmfkjolpipiffmcnipnpjilkjnmd}{Television Fanatic}}              & 139021    & \CIRCLE & \CIRCLE & \CIRCLE & \CIRCLE &         & \CIRCLE & \CIRCLE \\
    \texttt{\href{https://chrome.google.com/webstore/detail/biggamez-ads/aoeceebmempjbabimmnfkeeioccbjkea}{BigGamez ads}}                         & 47,394    &         & \CIRCLE &         &         &         &         & \CIRCLE \\
    \texttt{\href{https://chrome.google.com/webstore/detail/streamingtime/befmofeikcemiffcjbnphelmcamaommo/related}{StreamingTime}}               & 47,376    &         & \CIRCLE & \CIRCLE &         &         &         & \CIRCLE \\
    \texttt{\href{https://chrome.google.com/webstore/detail/convertowiz/ffmnloealihfajgcecbndhaehbhcdelf}{ConvertoWiz}}                           & 44,893    & \CIRCLE &         &         &         & \CIRCLE &         &         \\
    \texttt{\href{https://chrome.google.com/webstore/detail/utilitool-default-search/mhkgomffpgdlpgmggghambiakjjnffei}{Utilitool Search}}         & 34,272    &         & \CIRCLE & \CIRCLE &         &         &         & \CIRCLE \\
    \texttt{\href{https://chrome.google.com/webstore/detail/stream-itonline-default-s/gkcoogakgnilhknmndfahgciobihfmjj}{Stream-It.online Search}} & 30,887    &         & \CIRCLE &         &         &         &         & \CIRCLE \\
    \texttt{\href{https://chrome.google.com/webstore/detail/web-search/mmhgkibkmjbjobpldeceddgneehmidfk?hl=en}{Web Search}}                       & 24,992    & \CIRCLE & \CIRCLE &         &         & \CIRCLE &         &         \\
    \texttt{\href{https://chrome.google.com/webstore/detail/speedtest3-ads/gfpioeglfjecbkeeomdidlndcagpbmjf}{SpeedTest3 ads}}                     & 331       &         & \CIRCLE &         &         &         &         & \CIRCLE \\
    \bottomrule
  \end{tabular}
  \addtolength{\tabcolsep}{2pt}
  \begin{tablenotes}[online]
    \item[{[1]}] Display notifications.
    \item[{[2]}] Read and change all your data on a limited subset of websites.
    \item[{[3]}] Manage your apps, extensions, and themes.
    \item[{[4]}] Replace the page you see when opening a new tab.
    \item[{[5]}] Read your browsing history.
    \item[{[6]}] Manage your downloads.
    \item[{[7]}] Read and change all your data on the websites you visit.
    \end{tablenotes}
  \end{threeparttable}
  \vspace{-1em}
\end{table}

We did not go so far as to install these Chrome extensions, but we have little
doubt that their sole purpose is to record and sell data to advertisers, or to
steal users information when they visit other websites. This observation is
noteworthy in light of the claim made in~\cite{thomas2015extensions}: ``We
detect and report 192 deceptive Chrome extensions\ldots all have since been
disabled. Our techniques for catching these extensions are now used by Google to
scan new and updated extensions.'' Clearly, this method has not been
particularly successful, as these extensions still appear to be a preferred
method of exploiting end users. Successful maintenance of browser extensions
will likely benefit from coordinated efforts of multiple
viewpoints~\cite{dekoven2017malicious} using a variety of
techniques~\cite{kapravelos2014hulk}.

\begin{figure}
  \centering
  \frame{\includegraphics[width=1.0\columnwidth]{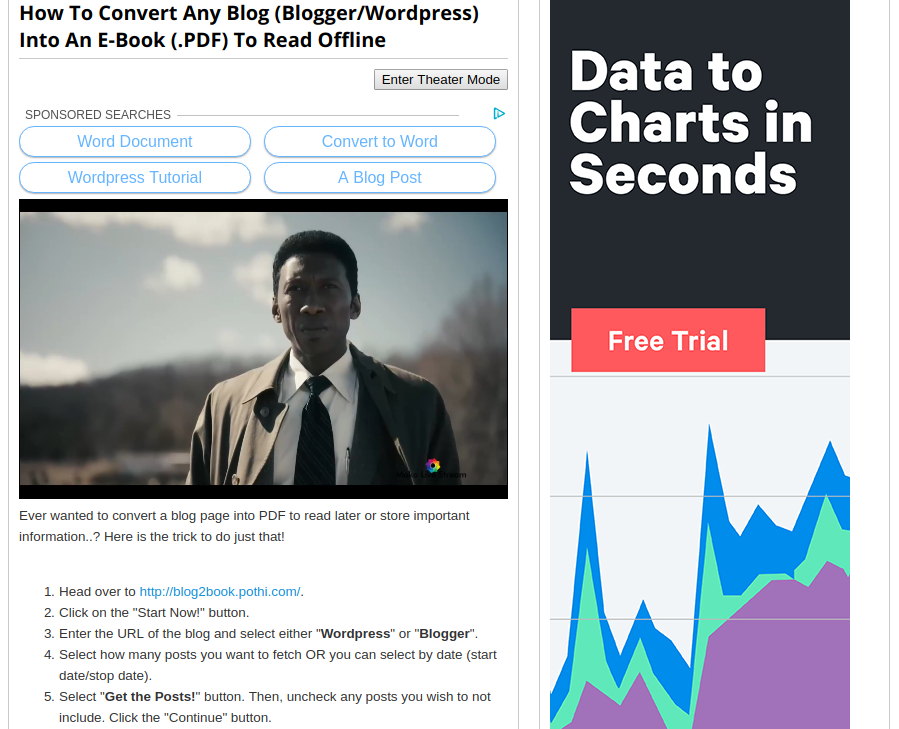}}
  \caption{Embedded live stream amid innocuous content on \texttt{mokoshalb.com}.
    On some sites, when a stream is over, the video element is removed,
    leaving only innocuous content.}
  \label{fig:mokoshalb}
  \vspace{-1em}
\end{figure}

\para{Hidden/Disguised Streams.} Several channel providers employed techniques
to hide or disguise the URLs of their live video streams. In many cases, it is
to the channel provider's benefit that their site appears innocuous to someone
casually browsing search results in order to not draw attention to their illegal
activity. We found two notable examples of this technique. First, is
\texttt{buffstreamz.com}, whose homepage appears to contain links to innocuous
sports news, but serves streams through hidden paths (e.g.,
\texttt{buffstreamz.com/watch/nba-1.php}) that redirect back to the homepage
whenever the stream is not live. Second, is \texttt{mokoshalb.com}, whose
homepage does not contain any content, but contains meta information suggesting
the website contains tutorials and tech tricks. \texttt{mokoshalb.com} takes
this one step further by also disguising the URL and the content of the video
stream page itself (Figure~\ref{fig:mokoshalb}). Other example stream URL paths
are shown in Table~\ref{tab:mokoshalb}.

\begin{table}
  \scriptsize
  \centering
  \caption{Example URL paths from \texttt{mokoshalb.com}}\label{tab:mokoshalb}
  \begin{tabular}{l}
    \toprule
    \textbf{URL Path} \\
    \midrule
    \texttt{/2017/04/top-12-google-search-tricks.html}             \\
    \texttt{/2017/04/seo-strategies-for-internet-marketing.html}   \\
    % \texttt{/2017/04/how-to-convert-any-blog.html}                 \\
    % \texttt{/2017/04/top-5-free-seo-testing-tools-for-new.html}    \\
    \texttt{/2016/06/9-things-to-consider-before-launching.html}   \\
    % \texttt{/2016/06/comparing-features-of-intels-core-i3-i5.html} \\
    \texttt{/2016/06/advantages-and-disadvantages-of-free.html}    \\
    % \texttt{/2016/06/blogger-vs-wordpress.html}                    \\
    % \texttt{/2016/06/8-tested-working-tips-to-be-better.html}      \\
    \texttt{/2016/06/how-to-make-money-with-affiliate.html}        \\
    \bottomrule
  \end{tabular}
  \vspace{-1em}
\end{table}

\para{Legitimate Affiliate Links} Some sites took an indirect approach to
monetizing ads by forcing users to follow redirects to Amazon product pages via
their own affiliate links in order to watch the provided streams. We found that
\texttt{firstrowonly.eu} occasionally redirects users to such Amazon product
pages with their own affiliate links the first time a user clicks anywhere on
the page, and that clicking ``watch now'' ads on the page occasionally redirects
to the Hulu trial page with an associated affiliate link.

\para{Obfuscation} Another common theme we found when visiting these websites
was that they would go to varying levels of effort to disguise their habits from
a casual investigator. For example, We observed that
\texttt{watchkobe.info/espn.php} modified its behavior if the Chrome Developer
Tools tab was opened, immediately hiding all ads as well as some overlay
\texttt{iframe}s that redirect to external pages. Additionally, we often
encountered URLs that were hidden from any casual scans through URI encoding or
Base64 encoding the URLs in the page source, and then having \texttt{onClick}
methods decode these URLs before navigating to them.

\begin{figure}
  \centering
  \includegraphics[width=1.\columnwidth]{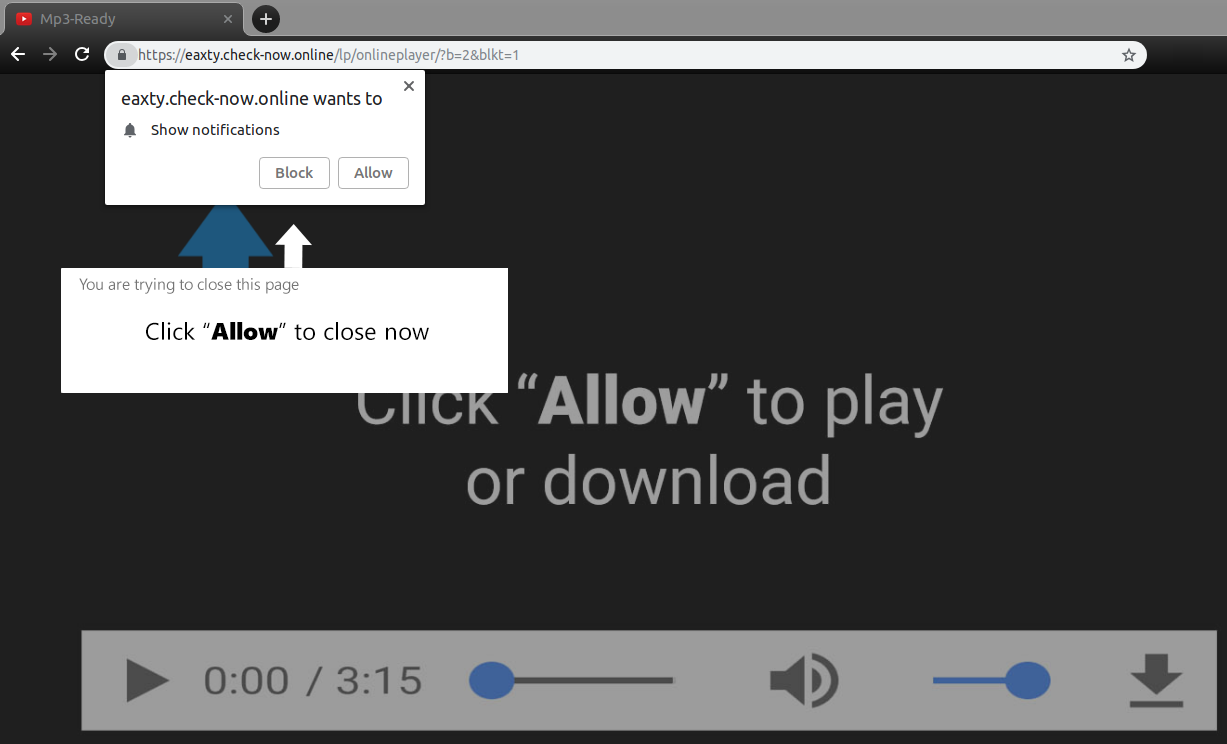}
  \caption{Techniques for gaining browser permissions}
  \label{fig:browser_notif}
  \vspace{-1em}
\end{figure}

\para{Browser Alerts} A particularly nasty technique which we frequently
encountered involved sites tricking users into believing they could only play
video, or only close the window, if they first accepted a browser notification,
via arrows on the page that pointed at where the browser notification would
appear. An example of this is shown in Figure~\ref{fig:browser_notif}.

\section{Implications}
\label{sec:implications}

As seen in Section~\ref{sec:analysis}, tracking on illegal live streaming
websites is widespread via third party HTTP requests, cookies, and device
fingerprinting. We want to draw attention to the significance of such tracking.
Though these free streaming websites are able to collect device fingerprints and
cookies, these sites rarely request access to the actual identity of users, and
thus, such tracking may seem inconsequential. Accordingly, we direct readers to
\cite{nikiforakis2013cookieless,acar2014web}, to remind them that linking
fingerprints to identities is often possible post-collection. For example, these
websites could sell these unique profiles to any online entity that does have
access to user identity, such as an ISP, an email operator, or a social network.
These entities would then be capable of comparing this known user identity and
their own collection of fingerprints with the fingerprints obtained from these
illicit sites, thus obtaining a list of people who view illegal streams. There
are many scenarios in which a legitimate company could benefit from such a
list---it could be useful for targeted ads about low-cost TV services or sports
betting websites. A more salacious example involves an ISP/cable company
desiring to throttle the Internet speed of any device suspected of violating the
exclusive broadcast rights of certain sports content. Such a list could also be
sold to more malicious actors and used as part of a ransomware scheme in which
users have to pay some entity under threat of being exposed for viewing illegal
content. These toy examples encompasses only a fraction of the different ways in
which such fingerprints could be used for harm, and highlight that the tracking
mechanisms detected in this work are far from harmless.

\section{Recommendations}
\label{sec:recommendations}

Our experience studying free illegal live streaming sites has led us to notice
several common patterns which are used to deceive users. Here, we make
recommendations for addressing these issues. Our purpose in presenting these
recommendations is to highlight them as potential areas for future work, and to
encourage discussion around these issues. For the recommendations we present, we
do not empirically evaluate their efficacy, nor do we claim that they are
infallible.

The most obvious pattern we found was that it was common for aggregators and
channel providers to leverage transparent \texttt{iframe}s which covered the
majority of the page, such that a user is essentially coerced to first click on
the \texttt{iframe} before interacting with the page. This technique, which is
similar to clickjacking, has been a longstanding problem with a variety of
proposed solutions~\cite{patil2016preventing, rydstedt2010busting,
shamsi2014clicksafe}. Building in more robust clickjacking protections into
common browsers may help raise awareness and provide a baseline level of
security for everyday users.

These deceptive overlays often prompt users to install malicious browser
extensions which masquerade as legitimate, but instead inject advertisements or
collect personal data. At the time of this writing, the Chrome Webstore does not
show an extension's version history and requested permissions up front on the
homepage of the extension. Furthermore, we found that different versions of the
same extension may display a different number of downloads and different
reviews. Different version can be directly linked to, but are not discoverable
via search or from the latest extension page. Chrome only shows the permission
requested by an extension after a user has consciously decided to click the
``Add to Chrome'' button. In contrast, Firefox Add-ons show both the permissions
an add-on requires, and allows a user to browse the version history before
clicking ``Add to Firefox''. With Chrome, a user decides to add an extension,
and then is shown the permissions and given an opportunity to change her mind.
With Firefox, the user has all the information up front before needing to make
that initial decision, and is also given an opportunity to change her mind.

In our experience, malicious browser extensions represented the most
user-hostile behavior we found on these websites. Notably, we were not prompted
to download or install any binaries or desktop applications. All of the prompts
we encountered for malicious software took the form of browser extensions.
Although significant progress has been made in recent years to fight malicious
extensions~\cite{jagpal2015trends}, it is clear that malicious extensions are
still a prevalent attack vector, and further work in automatic detection of
potentially malicious extensions would benefit users.

\section{Conclusion}
\label{sec:conclusion}

In this paper, we extend and update the body of work investigating illegal
streams. We discover that the streaming ecosystem has continued to expand, with
Reddit emerging as one of the largest aggregators of illegal streams. We collect
links to hundreds of thousands of streams over the period of a month, and crawl
these websites using OpenWPM, which leads us to discover that these websites go
to significant lengths to track users---and that they do so in a much more
comprehensive and unavoidable manner compared to legal sports streaming
websites. Further, our personal observations of dozens of channel providers lead
us to the conclusion that there is still a large range of options for monetizing
page visits beyond traditional ads for criminals willing to exploit users. We
find that deceptive ads and full-page overlay redirects are commonly used, and
that techniques to deceive users into installing malicious extensions are
commonplace. Further, the frequency with which we observe tracking and
distribution of user specific information implies that these sites also
monetarily benefit from from tracking users, in addition to serving
advertisements. We caution that this might indicate at least some of these sites
are able to sell this tracking data to third parties, where these third parties
could be advertisers or more malicious entities.

We believe that this piece has shown that illegal streaming represents a
uniquely accessible view into misbehaving websites on the Internet. These sites
are by definition criminal enterprises, but require substantial audiences in
order to profit---as a result, they are easy to locate, and make little attempt
to hide from security researchers. These sites serve as an excellent case study
of modern techniques used to profit off of users, whether that be via deceptive
ads, abuse of affiliate programs, user tracking, or distributing malware. We
encourage future security research to take advantage of the illegal streaming
ecosystem as a resource for studying how cybercriminals monetize page views, and
a tool for measuring modern tracking techniques employed on the web.

\para{Acknowledgements}
We would like to thank Zakir Durumeric and Dima Kogan for their advice and
guidance throughout this project. We would also like to acknowledge the
constructive feedback offered by the students of CS356 at Stanford University.

{\footnotesize \bibliographystyle{acm}
  \bibliography{main}
}

\end{document}